\documentclass{elsart}
\usepackage{graphics}
\usepackage{epsfig}
\usepackage{amsfonts}
\usepackage{amssymb}
\usepackage{amsmath}
\usepackage{cite}
\begin{document}
\begin{frontmatter}
\title{The $\boldsymbol{pd\rightarrow\hspace{0mm}^3}$He$\hspace{1mm}\boldsymbol{\eta\pi^0}$ reaction at $\boldsymbol{T_p = 1450}$~MeV}
%\title{The $p d \rightarrow ^3$He$\,\eta\,\pi^0$ reaction at $T_p$ = 1450 MeV}

\author[Uppsala]{K.~Sch\"onning\corauthref{cor1}},
\ead{karin.schonning@fysast.uu.se}
\ead[url]{www3.tsl.uu.se/$\sim$schonnin}\corauth[cor1]{Corresponding author.}
\ead{karin.schonning@fysast.uu.se}
\author[Stockholm]{Chr.~Bargholtz},
\author[Tuebingen]{M.~Bashkanov},
\author[Warsaw]{M.~Ber{\l}owski},
\author[Dubna]{D.~Bogoslawsky},
\author[Uppsala]{H.~Cal\'en},
\author[Tuebingen]{H.~Clement},
\author[Hamburg]{L.~Demir\"ors},
\author[TSL]{C.~Ekstr\"om},
\author[Uppsala]{K.~Fransson},
\author[Stockholm]{L.~Ger\'en},
\author[Uppsala]{L.~Gustafsson},
\author[Uppsala]{B.~H\"oistad},
\author[Dubna]{G.~Ivanov},
\author[Uppsala]{M.~Jacewicz},
\author[Dubna]{E.~Jiganov},
\author[Uppsala]{T.~Johansson},
\author[Uppsala]{S.~Keleta},
\author[Tuebingen]{O.~Khakimova},
\author[Tuebingen]{F.~Kren},
\author[Uppsala]{S.~Kullander},
\author[Uppsala]{A.~Kup\'s\'c},
\author[Novosibirsk]{A.~Kuzmin},
\author[Stockholm]{K.~Lindberg},
\author[Uppsala]{P.~Marciniewski},
\author[Dubna]{B.~Morosov},
\author[Juelich]{W.~Oelert},
\author[Hamburg]{C.~Pauly},
\author[Uppsala]{H.~Petr\'en},
\author[Dubna]{Y.~Petukhov},
\author[Dubna]{A.~Povtorejko},
\author[Hamburg]{W.~Scobel},
\author[Moscow]{R.~Shafigullin},
\author[Novosibirsk]{B.~Shwartz},
\author[Tuebingen]{T.~Skorodko},
\author[Moscow2]{V.~Sopov},
\author[Warsaw]{J.~Stepaniak},
\author[Stockholm]{P.-E.~Tegn\'er},
\author[Uppsala]{P.~Th\"orngren Engblom},
\author[Dubna]{V.~Tikhomirov},
\author[Warsaw2]{A.~Turowiecki},
\author[Tuebingen]{G.~J.~Wagner},
\author[UCL]{C.~Wilkin},
\author[Juelich]{M.~Wolke},
\author[Lodz]{J.~Zabierowski},
\author[Stockholm]{I.~Zartova} and
\author[Uppsala]{J.~Z{\l}oma\'nczuk}
\collaboration{The CELSIUS/WASA Collaboration} 

%\author[Uppsala]{B.~H\"{o}istad} %
% \author[Where]{Authors to be decided},
%\author[UCL]{C.~Wilkin}

\address[Uppsala]{Department of Physics and Astronomy, Uppsala University,
Box 535, S-751 21 Uppsala, Sweden}
\address[Stockholm]{Department of Physics, Stockholm University, S-101 91 Stockholm, Sweden}
\address[Tuebingen]{Physikalisches Institut der Universit\"at T\"ubingen, D-720 76 T\"ubingen, Germany}
\address[Warsaw]{So{\l}tan Institute of Nuclear~Studies, PL-006 81 Warsaw, Poland}
\address[Dubna]{Joint Institute for Nuclear Research, Dubna, 101 000 Moscow, Russia}
\address[Hamburg]{Institut f\"ur Experimentalphysik, Universit\"at Hamburg, D-227 61 Hamburg, Germany}
\address[TSL]{The Svedberg Laboratory, S-751 21 Uppsala, Sweden}
\address[Novosibirsk]{Budker Institute of Nuclear Physics, Novosibirsk 630 090, Russia}
\address[Juelich]{Institut f\"ur Kernphysik, Forschungszentrum J\"ulich GmbH, D-524 25 J\"ulich, Germany}
\address[Moscow]{Moscow Engineering Physics Institute, Moscow, Russia}
\address[Moscow2]{Institute of Theoretical and Experimental Physics, Moscow, Russia}
\address[Warsaw2]{Institute of Experimental Physics, PL-006 81 Warsaw, Poland}
%\address[Where]{Lots of addresses}
\address[UCL]{Physics and Astronomy Department, UCL, London, WC1E 6BT, UK}
\address[Lodz]{Soltan Institute of Nuclear Studies, PL-901 37 Lodz, Poland}
\begin{abstract}
The cross section for the $pd\to \,^{3}\textrm{He}\,\eta\,\pi^0$ reaction
has been measured at a beam energy of 1450~MeV using the WASA detector
at the CELSIUS storage ring and detecting one $^3$He and four photons from the decays of the two mesons. The data indicate that the production mechanism
involves the formation of the $\Delta(1232)$ isobar. Although the beam
energy does not allow the full peak of this resonance to be seen, the
invariant masses of all three pairs of final particles are well reproduced
by a phase space Monte Carlo simulation weighted with the $p$-wave factor
of the square of the $\pi^0$ momentum in the $^3$He$\,\pi^0$ system.
\end{abstract}

\begin{keyword}
$\eta$, $\pi^0$ meson production, invariant mass distribution, $\Delta$ resonance;%
\PACS 25.40.Ve %Other reactions above meson production thresholds (energies > 400 MeV)%
\sep 14.40.Be %Other mesons with S=C=0, mass < 2.5 GeV%
\sep 13.75.-n %hadron induced low energy interactions
\end{keyword}
\end{frontmatter}
%
%%%%%%%%%%%%%%%%%%%%%%%%%%%%%%%%%%%%%%%%%%%%%%%%%%%%%%%%%%%%%%%%%
%
\newpage

The $pd\to\,^3$He$\,X^0$ reaction has long been used to study the
production of neutral mesons or mesonic systems. Missing-mass
experiments carried out near the production thresholds have clearly
identified peaks corresponding to $X^0 = \omega$, $\eta'$, and
$\phi$~\cite{WUR95,WUR96}. Of particular interest are the data on the
production of the $\eta$ meson~\cite{BER88,MAY96,MER07,SMY07}, which
show a threshold enhancement that might indicate the formation of a
quasi-bound $\eta\,^3$He nuclear state~\cite{WIL93}. Evidence in
favour of this hypothesis is to be found also in the coherent $\eta$
photoproduction from $^3$He,
$\gamma\,^3$He$\to\eta\,^3$He~\cite{PFE04}.

However, exclusive measurements of a production process often yield important
additional information. The study of $pd\to\,^3$He$\,K^+K^-$ showed
that the $\phi$ mesons produced and decaying into $K^+K^-$ are
strongly polarised with respect to the incident proton
direction~\cite{BEL07}. In contrast, the $\omega$ mesons detected
through the measurement of $pd\to\,^3$He$\,\pi^+\pi^-\pi^0$ have very
low polarisation~\cite{SCH08}. This difference is in marked contrast
to the Okubo-Zweig-Iizuka rule~\cite{OZI}, which would suggest rather
that the polarisations of these two vector mesons should be similar.

The most quoted data on the $pd\to\,^3$He$\,X^0$ reaction are
connected with the ABC effect, where a strong and sharp enhancement
of the missing mass $X^0$ spectrum is seen a little above the
threshold of two pions~\cite{ABA60}. The effect might be connected
with the production of two $\Delta(1232)$ isobars or with the
sequential decay of the Roper $N^*(1440)$ resonance. However, the
full rich structure could only be made accessible through exclusive
measurements, such as those carried out recently for
$pd\to\,^3$He$\pi^0\pi^0$ and $pd\to\,^3$He$\pi^+\pi^-$~\cite{BAS06}.
It is interesting to see if any similar ABC effect were to be found
in the production of other pairs of pseudoscalar mesons, such as
$\eta\,\pi^0$. In this case an exclusive measurement would be
required in order to identify the reaction against the much larger
background arising from multipion production.

Many important results have appeared recently on the photoproduction
of the $\pi^0\,\eta$ system. The data from hydrogen~\cite{KAS09} have
been interpreted in terms of a dominant cascade decay of the $D_{33}$
$\Delta(1700)$ isobar through the $s$-wave $\Delta(1700)\to
\eta\Delta(1232)$ followed by the $p$-wave $\Delta(1232)\to \pi^0
p$~\cite{DOR06}. The evidence for the importance of the
$\Delta(1232)$ is clear from the invariant mass distribution, though
some signal of the interaction of the $\eta$ with the observed proton
through the $N^*(1535)$ is also apparent~\cite{KAS09}. The coherent
photoproduction of $\pi^0\,\eta$ pairs in $\gamma d\to \eta\,\pi^0d$
has also been observed~\cite{JAE09}. The positive signal of the
similar reaction on $^3$He raises the tantalising possibility of
using the $\gamma\,^3\textrm{He}\to \eta\,\pi^0\,^3\textrm{He}$
reaction to study also the final state interaction (\textit{fsi}) of
the $\eta$ with the $^3$He~\cite{KRU09}. The competition between the
$\eta\,^3$He and $^3$He$\,\pi^0$ interactions would, of course, also
be equally relevant if the system were produced in proton-deuteron
collisions.

Measurements of the $pd\to \,^{3}\textrm{He}\,\eta\,\pi^0$ reaction
were carried out at the CELSIUS storage ring of the The Svedberg Laboratory in Uppsala, Sweden,
using the WASA detector~\cite{Zabierowski}. The circulating
proton beam of energy 1450~MeV was incident on a deuterium pellet
target~\cite{Ekstrom,Nordhage}. The $^3$He ejectiles were measured in
the WASA forward detector (FD) \cite{CAL96}, which covered laboratory polar angles
from $3^{\circ}$ to $18^{\circ}$. This corresponds to 92\% of the
$^3$He phase space for $\eta \pi^0$ production at 1450~MeV. The lost
events are those where the $^3$He are emitted at small laboratory
angles such that they escape detection down the beam pipe.

The forward detector consists of a sector-like window counter
(FWC) for triggering, a proportional chamber for precise angular
information (FPC), a hodoscope (FTH) for triggering and off-line
particle identification, a range hodoscope (FRH) for energy
measurements, particle identification and triggering, and a veto
hodoscope (FVH) for triggering.

The $\eta$ and $\pi^0$ mesons were identified \textit{via} their
decay into $\gamma\gamma$ pairs, with these photons being measured in
the central detector (CD). Their energies and directions were
determined using the information from the Scintillating
Electromagnetic Calorimeter (SEC), which covers polar angles from
$20^{\circ}$ to $169^{\circ}$. The absence of a signal in the Plastic
Scintillating Barrel (PSB) indicated that the photons arose from the decay
of a neutral particle.

\begin{figure}[hbt]
\begin{center}
\includegraphics[width=0.8\textwidth]{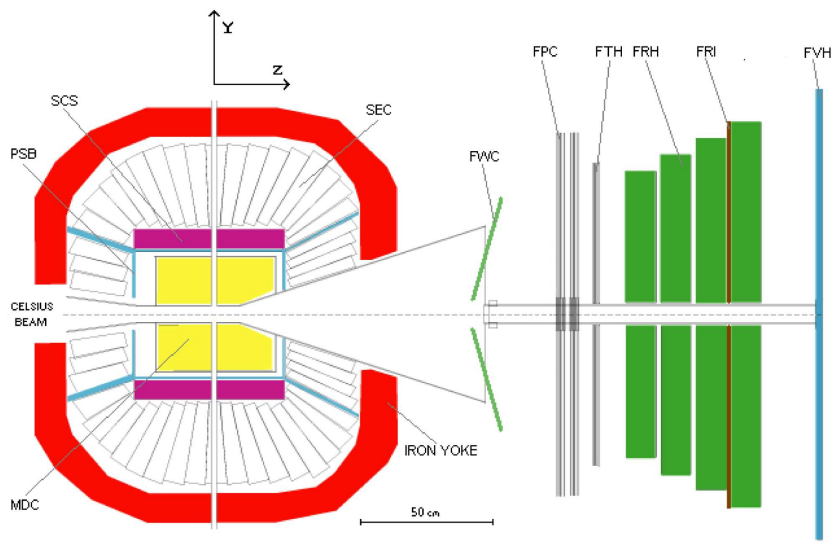}
\caption{(Colour online) Side view of the CELSIUS/WASA detector setup \cite{Zabierowski,CAL96}. The CELSIUS
beam pipe runs horizontally and the target pellets are injected
downwards through the vertical pipe.} \label{fig:wasa4pi}
\end{center}
\end{figure}

A schematic overview of the WASA detector setup is shown in
Fig.~\ref{fig:wasa4pi}.%\vspace{5mm}

The hardware $^3$He trigger selected events where there was a hit
with a high energy deposit in the FWC and an overlapping hit in
either the FTH or the FRH. The $^3$He were identified in the FD using
the $\Delta E\!-\!E$ method, as described in detail in
Refs.~\cite{karin1,karinthesis}.

In the data analysis we considered only those events where the $\eta$
meson decayed into two photons (BR = 39.3\%), and therefore selected
events with one $^3$He plus four photons. Furthermore, one
$\gamma\gamma$ combination was required to have an invariant mass
close to that of the $\pi^0$, $|IM(\gamma\gamma)-m_{\pi^0}| <
45$~MeV/$c^2$. The two remaining photons must have an opening angle
$\theta_{\gamma\gamma}^{\eta}> 70^{\mathrm{o}}$, motivated by Monte Carlo simulations of the reaction, and an invariant mass
larger than 460~MeV/$c^2$. In addition, the overall missing mass
should be small, $MM(^3$He$\,4\gamma)< 100~\rm{MeV}/c^2$. Finally,
all events with two $\pi^0$ candidates, \textit{i.e.}, where two
$\gamma\gamma$ combinations satisfied $|IM(\gamma\gamma)-m_{\pi^0}| <
45$~MeV/$c^2$, were rejected. This reduced the background
contribution from $2\pi^0$ production by almost an order of
magnitude. These selection criteria, when applied on phase space produced $pd \rightarrow ^3$He$\,\eta\pi^0,\,\eta \rightarrow \gamma\gamma$, lead to an acceptance of 11.1\%.

The above cuts reduce the acceptances for $2\pi^0$ and $3\pi^0$
production to $\approx 0.1\%$ and $\approx 0.2\%$, respectively.
However, since their cross sections are so much larger than that for
the $\eta\,\pi^0$ channel, and only 39.3\% of the $\eta$ mesons decay
into $\gamma\gamma$, a significant background from multipion
production will remain.

\begin{figure}[hbt]
\begin{center}
\includegraphics[width=0.8\textwidth]{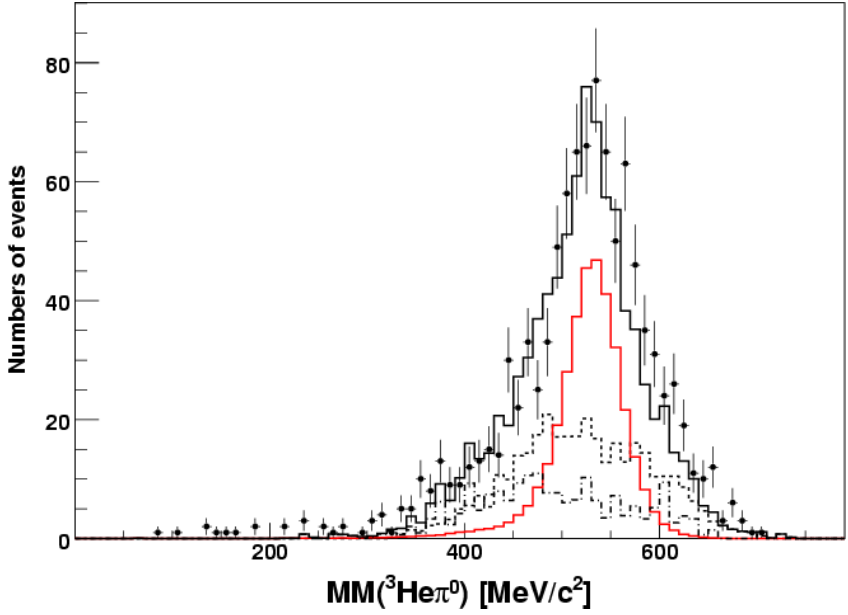}
\caption{(Colour online) The missing mass of the $^3$He$\,\pi^0$ system for all
events fulfilling the selection criteria given in the text. The
dash-dotted line represents simulated $p d \to \, ^3$He$\,2\pi^0$
events, the dotted line $p d \to \, ^3$He$\,3\pi^0$, and the red line
$p d \to \, ^3$He$\,\eta\,\pi^0$. The weights of these three
contributions have been adjusted so that their sum (solid black line)
reproduces well the experimental data (points).}
\label{fig:mm_3hepi0}
\end{center}
\end{figure}

The $pd\to\, ^3$He$\,\eta\,\pi^0$ events are identified by the peak
at the $\eta$ position that appears in the $^3$He$\,\pi^0$ missing
mass spectrum shown in Fig.~\ref{fig:mm_3hepi0}. The points are
experimental data that satisfy the selection criteria. Phase space
simulations are shown of  $p d \to \, ^3$He$\,2\pi^0$ (dash-dotted
line), $p d \to \, ^3$He$\,3\pi^0$ (dotted line), and $p d \to \,
^3$He$\,\eta\,\pi^0$ (solid red line). The three contributions are
normalised such that their sum (solid black line) gives the best fit
to the experimental data. The $2\pi^0$ and $3\pi^0$ distributions are
then roughly consistent with the cross sections obtained in
Ref.~\cite{karinthesis}. The $\eta\,\pi^0$ distribution normalised in
this way contains $375\pm 35$ events, where the quoted error is
systematic, mainly arising from the ambiguity in the background subtraction.

The number of $\eta\,\pi^0$ candidates is corrected for acceptance, taking the $\eta \to \gamma\gamma$ branching ratio into account, and then divided by
the integrated luminosity, which was determined as described in
Ref.~\cite{karinomega}, in order to obtain the total cross section.
This procedure gave a value of $\sigma_{\rm tot} = 22.6 \pm 1.5 \pm
2.1 \pm 14\%$.  The first error is statistic and the second systematic, coming from
uncertainties in the number of $\eta\,\pi^0$ events and the
acceptance estimation. The third
reflects the uncertainty in the normalisation, in which effects from
both the luminosity (12\%) and time-overlapping events ($<8\%$) are
included, being added quadratically.

In the $pd \to \, ^3$He$\,\eta\,\pi^0$ reaction there are potentially
three important final state interactions, which have been investigated by
constructing the invariant mass distributions for the $\eta\,\pi^0$,
$^3$He$\,\pi^0$, and $^3$He$\,\eta$ systems. For this purpose we take
all the events in Fig.~\ref{fig:mm_3hepi0} that lie within the
interval $490~\rm{MeV}/c^2 < MM(^3\rm{He}\,\pi^0) <
580~\rm{MeV}/c^2$. This asymmetric choice is motivated by the fact
that the $\eta$ peak is shifted towards lower masses in both the
Monte Carlo simulation as well as in the experimental data. Within
this mass interval there are $\approx 335~\eta\,\pi^0$ candidates, with a
signal-to-background ratio of 1.7.

\begin{figure}[hbt]
\begin{center}
\includegraphics[width=0.8\textwidth]{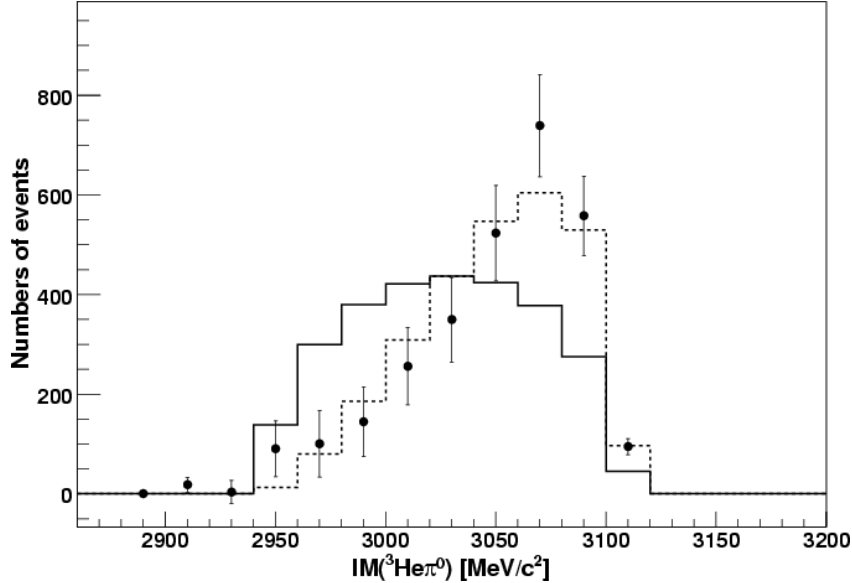}
\caption{The invariant mass of the $^3$He$\,\pi^0$ system for all
events fulfilling the selection criteria given in the text and, in
addition, $490 < MM(^3$He$\,\pi^0) < 580$~MeV/$c^2$. The background
has been subtracted and the distribution corrected for acceptance.
The error bars represent only the statistical uncertainties. The
solid lines show a phase space simulation $^3$He$\,\eta\,\pi^0$
events. The dotted histogram shows phase space events weighted with the square of the $\pi^0$ momentum in the
$^3$He$\,\pi^0$ rest system.} \label{fig:im_pihe}
\end{center}
\end{figure}

Figure \ref{fig:im_pihe} shows the invariant mass of the
$^3$He$\,\pi^0$ system, where the background, obtained from simulated
$2\pi^0$ and $3\pi^0$ data has been subtracted from each bin. The
remaining numbers of events have been corrected for acceptance, also
estimated bin by bin. The results are shown as points with error bars
that represent the statistical uncertainties. In addition to these
there is a systematic uncertainty in each bin of less than 10\% due
to the background subtraction and acceptance
estimation. The solid line shows a phase space simulation of
$pd\to\,^3$He$\,\eta\,\pi^0$ events. The experimental data peak
slightly below 3100~MeV/$c^2$, which is approximately equal to
$2m_p+M_{\Delta(1232)}$ and points towards an involvement of the
$\Delta(1232)$ isobar in the production process. At this energy, the
full $\Delta$ peak is not covered and the data are primarily
sensitive to the $p$-wave rise towards the resonance position. To
simulate this effect, Monte Carlo events have been weighted with
$k^2$, the square of the momentum of the $\pi^0$ in the
$^3$He$\,\pi^0$ rest frame. The resulting distribution is shown in
Fig.~\ref{fig:im_pihe} by the dotted histogram, where the
normalisation is to the total number of events. This model
reproduces well the shape of the data. 
%Any possible deviation at low energies might correspond to a small contribution from $s$-wave production or to an improperly subtracted background.

\begin{figure}[hbt]
\begin{center}
\includegraphics[width=0.8\textwidth]{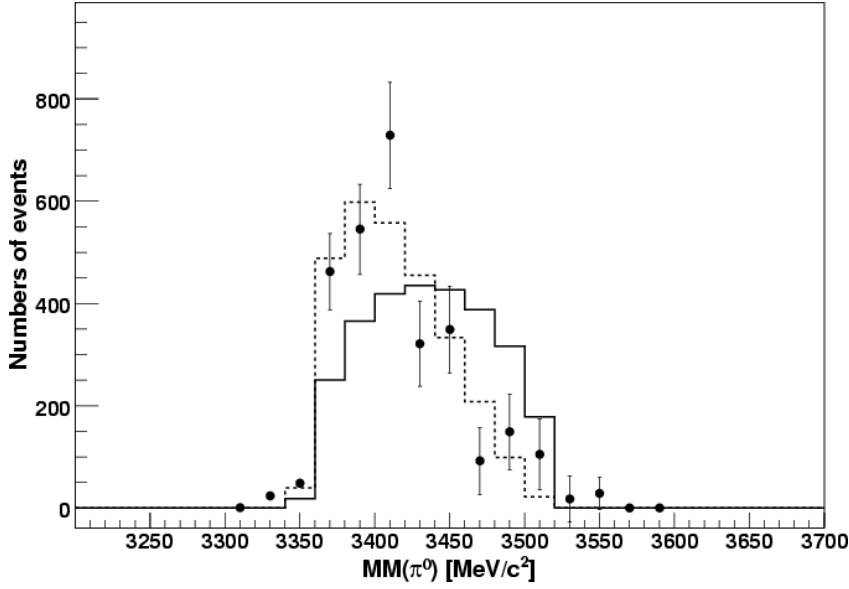}
\caption{The missing mass of the $\pi^0$ for all events
fulfilling the selection criteria given in the text and, in addition,
$490 < MM(^3$He$\,\pi^0)<580$~MeV/$c^2$. This distribution is equivalent to that of the
invariant mass of the $^3$He$\,\eta$ system. The solid lines represent
phase space $\eta\,^3$He Monte Carlo data and the dotted ones the
same but weighted by the square of the $\pi^0$ momentum in the
$^3$He$\,\pi^0$ rest system.} \label{fig:im_ethe}
\end{center}
\end{figure}

Instead of studying the invariant mass of the $^3$He$\,\eta$ system,
it is in practice more reliable to construct the missing mass of
the $\pi^0$. This is because the electromagnetic calorimeter was
calibrated using the neutral pions decaying into $\gamma\gamma$ so
that it is in this region more precise than for $\eta$ decay. The basic
procedure for obtaining the distribution is similar to that for the
$^3$He$\,\pi^0$ invariant mass. After subtracting the background, the
data were corrected for acceptance and the result is shown in
Fig.~\ref{fig:im_ethe}. Compared to the broadly semi-circular form of
the phase space distribution, the experimental data show a
peaking towards low missing masses. At first sight this might be
interpreted as being due to a $^3$He$\,\eta$ final state interaction,
which is very strong and attractive near the kinematic threshold.
However, the dotted histogram, again showing phase space simulations
weighted by the square of the $\pi^0$ momentum in the $^3$He$\pi^0$
rest frame, strongly suggests that this also could be an effect of
the $p$-wave interaction between the $\pi^0$ and the $^3$He.

\begin{figure}[hbt]
\begin{center}
\includegraphics[width=0.8\textwidth]{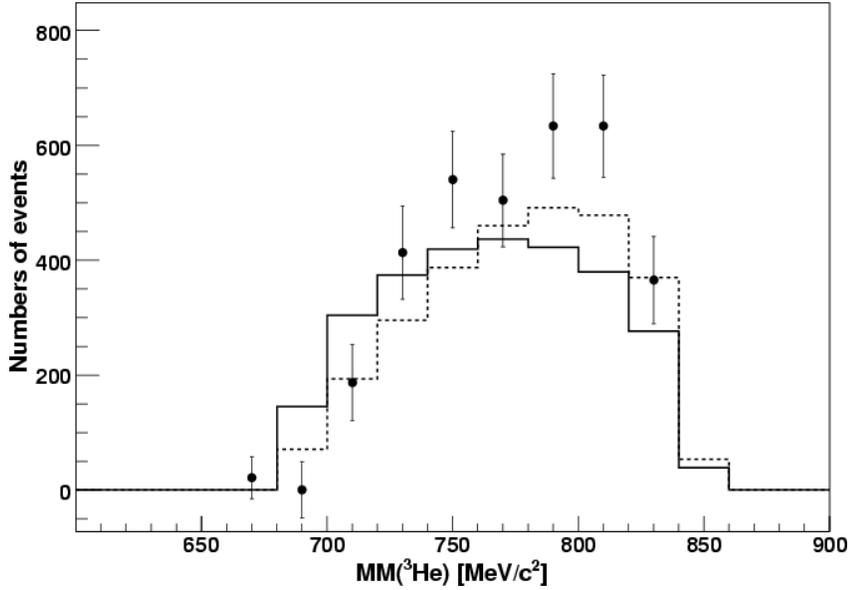}
\caption{The missing mass of the $^3$He for all events
fulfilling the selection criteria given in the text and when 490 MeV
$< MM(^3$He$\,\pi^0) <$ 580 MeV.
This distribution is equivalent to that of the
invariant mass of the  $\eta\,\pi^0$ system,
The solid curve is a Monte Carlo
simulation of phase space production, while the dotted
curve represents these events weighted by the square of the
$\pi^0$ momentum in the $^3$He$\,\pi^0$ rest system.}
\label{fig:mm_3he}
\end{center}
\end{figure}

The best measurement of the $\eta\,\pi^0$ invariant mass is obtained
through the study of the $^3$He missing mass because the nucleus is
detected in the Forward Detector, which has a much better
resolution than the electromagnetic calorimeter. The background
subtracted and acceptance corrected results are shown in
Fig.~\ref{fig:mm_3he}. The deviations from phase space are not so
marked as in the cases that involved the $^3$He but even here the
small effects are fairly well reproduced by weighting the Monte Carlo
simulation with the $k^2$ factor. It is, of course, not surprising that
one sees no significant influence of the $a_0(980)$ scalar resonance since at
$T_p = 1450$~MeV the maximum $\eta\,\pi^0$ invariant mass that is accessible
is only about 850~MeV/$c^2$.

Although the $p$-wave \textit{ansatz} reproduces all three final
invariant mass distributions very economically through the
introduction of the $k^2$ factor, there is no sign of any $p$-wave
nature in the angular distributions where, within the limited
statistics, the data are fairly isotropic in the angle between the
proton and the $\pi^0$ in the overall c.m.\ frame. The same is true
for the angle of the $\pi^0$ in the $^3$He$\,\pi^0$ frame with
respect to either the incident proton or recoiling $\eta$. However,
in view of the large number of spin degrees of freedom it is hard to
draw conclusions from such isotropy.

Since WASA has a very large acceptance, the introduction of the $k^2$ factor
into the Monte Carlo has only a limited effect, changing the total acceptance from 11.1\% to 10.6\% which changes the value of the
total cross section for the reaction to $\sigma_{\rm tot} = 23.6 \pm 1.6 \pm
2.2 \pm 14\%$.

In summary, we have carried out measurements of the $pd\to
\,^{3}\textrm{He}\,\eta\,\pi^0$ reaction at a beam energy of
1450~MeV. Although the statistics are not sufficient to make a useful
Dalitz plot, the invariant mass distributions of all three final
pairs of particles are consistent with the $p$-wave influence that
might arise from the formation of the $\Delta(1232)$ in the
$^3$He$\,\pi^0$ system. This is very much in line with the
photoproduction data on hydrogen and deuterium obtained at higher
excess energies~\cite{KAS09,JAE09}. There is no sign of any
enhancement of the ABC type in the $\eta\,\pi^0$ mass distribution
and the angular distributions, which within large error bars are consistent with isotropy, are in marked contrast
to the very rich structure observed for
$pd\to\,^3$He$\,\pi\pi$~\cite{BAS06}.

It would be highly desirable to have a microscopic model for the
$pd\to \,^{3}\textrm{He}\,\eta\,\pi^0$ reaction. In particular it is
important to identify the dynamical origin of the $\eta$. Does it
come from a sequential decay of the $D_{33}$ $\Delta(1700)$ isobar,
as suggested for the photoproduction data~\cite{KAS09,DOR06}, or does
it arise from a two-step process such as $pn\to d\eta$ followed by
$dp\to \,^3$He$\,\pi^0$, where the $N^*(1535)$ plays a role?
Regarding the final state interactions, it seems already clear from
our results that data would have to be obtained at higher energy in
order to separate the different final state interactions and to have
a chance of investigating the formation of any $\eta\,^3$He
quasi-bound state. The data would then extend over the peak of the
$\Delta(1232)$ and thus allow firmer conclusions to be drawn.
Experiments of this type could be carried out by the WASA-at-COSY
collaboration~\cite{COSY}.

\vspace{1pt}We are grateful to the personnel at The Svedberg
Laboratory for their support during the course of the experiment.
This work was supported by the European Community under the
``Structuring the European Research Area'' Specific Programme
Research Infrastructures Action (Hadron Physics, contract number
RII3-cT-204-506078), and by the Swedish Research Council.

%
%%%%%%%%%%%%%%%%%%%%%%%%%%%%%%%%%%%%%%%%%%%%%%%%%%%%%%%%%%%%%%%%%
%

\end{document}